\documentclass{article}
\usepackage{amssymb}

\input{tcilatex}
\begin{document}

\title{\textbf{Second quantization of time and energy in Relativistic
Quantum Mechanics}}
\author{M. Bauer* and C.A. Aguill\'{o}n** \\
*Instituto de F\'{\i}sica, **Instituto de Ciencias Nucleares,\\
Universidad Nacional Aut\'{o}noma de M\'{e}xico\\
e-mail: bauer@fisica.unam.mx}
\maketitle

\begin{abstract}
Based on Lorentz invariance and Born reciprocity invariance, the canonical
quantization of Special Relativity (SR) has been shown to provide a unified
origin for the existence of Dirac's Hamiltonian and a self adjoint time
operator that circumvents Pauli's objection. As such, this approach restores
to Quantum Mechanics (QM) \ the treatment of space and time on an equivalent
footing as that of momentum and energy. Second quantization of the time
operator field follows step by step that of the Dirac Hamiltonian field. It
introduces the concept of time quanta, in a similar way to the energy quanta
in Quantum Field Theory (QFT). An early connection is found allready in
Feshbach's unified theory of nuclear reactions. Its possible relevance in
current developments such as Feshbach resonances in the fields of cold atom
systems, of Bose-Einstein condensates and in the problem of time in Quantum
Gravity is noted. \ \ \ \ \ \ .
\end{abstract}

\bigskip \textbf{"One must be prepared to follow up the consequences of
theory, and feel that one just has to accept the consequences no matter
where they lead" }\textit{\ }P.A.M. Dirac\footnote{%
Quoted by J. Polchinsky of UCSB in "23d Solvay Conference - The Quantum
Structure of Space and Time", World Scientific, 2007}

\section{Introduction}

Resolving the Problem of Time is still a central issue in Quantum Mecanics
and Quantm Gravity\cite{Muga,Anderson,Kuchar}. Pauli'objection\footnote{%
\ \ \ \ \ \textit{\ \ \ "In the older literature on quantum mechanics, we
often find the operator equation Ht-tH=i}$\hslash $\textit{. It is generally
not possible, however, to construct a Hermitian operator (e.g. as a function
of P and Q) which satisfies this equation. This is so because, from the C.R.
written above it follows that H possesses continously all eigenvalues from -}%
$\infty $\textit{\ to +}$\infty $\textit{, whereas on the other hand,
discrete eigenvalues of H can be present. We, therefore, conclude that the
introduction of an operator t is basically forbidden and the time t must
necessarily be considered an ordinary number (`c' number) in Quantum
Mechanics"}}\cite{Pauli}to the existence of a time operator canonically
conjugate to the Hamiltonian did set time to remain a parameter (a $c$%
-number) while space coordinates were represented by self-adjoint operators
\ ($q$-numbers)\cite{Dirac}, foregoing the equal footing of space and time
accorded by Special Relativity (SR), as well as questioning the existence
and interpretation of a time energy uncertainty relation\cite{Busch,Bauer}.

However the canonical quantization of SR\cite{Aguillon,Bauer2}, together
with Born's reciprocity principle\cite{Born,Freidel}, has now been shown to
provide a formal basis for both the Dirac Hamiltonian and the existence of a
self-adjoint "time operator" in Relativistic Quantum Mechanics (RQM). As the
generator of continuous momentum displacements (there is no gap in the
momentum spectrum) this time operator induces consequently a shift of energy
in both branches of the energy relativistic spectrum, circumventing Pauli's
objection \cite{Bauer2,Bauer1,Bauer3}. Its eigenspinors provide an
orthonormal basis alternative to the energy momentum spinor basis provided
by the Dirac Hamiltonian, and consequently a different representation. Then
both can be subjected to second quantization to show that a field can be
equivalently expressed as a function of spacetime coordinates $\psi (x^{\mu
})$ or of energy--momentum coordinates $\phi (p_{%
{\mu}%
}),$ the two formulations being related by the ordinary Fourier transform.\ 

The present paper explores whether this new basic element of RQM provides
additional insights on the dynamics of many particle systems in the
ocupation number representation of Quantum Field Theory (QFT). To quote:
"Physics does not depend on the choice of basis, but which is the most
convenient choice depends on the physics"\footnote{%
A. Zee, \textit{"Quantum Field Theory in a Nutshell"}, Princeton University
Press (2010)}. Section 2 reviews briefly the derivation of the time
operator. Section 3 presents the second quantization of the time operator
field that follows step by step that of the Dirac Hamiltonian field.
Interpretation and conclusions are included in Section 4.

\section{First quantization: configuration, momentum, time and energy
representations\protect\cite{Aguillon}}

Canonical quantization\ and factorization of the special relativity free
particle invariant \ $p^{\mu }p_{\mu }=\pi ^{2}:=(m_{0}c)^{2}$\ yields a
constraint that is satisfied by the linear equation:%
\begin{equation}
\lbrack \gamma ^{\nu }\hat{p}_{\nu }-m_{0}c]\left\vert \Psi \right\rangle =0
\end{equation}%
provided that in the Minkowski metric $\eta ^{\mu \nu }=diag(1,-1,-1,-1)$
the following anticommutation and commutation relations are satisfied:%
\begin{equation}
\left\{ \gamma ^{\mu },\gamma ^{\nu }\right\} =2\eta ^{\mu \nu }\mathbf{I}%
\text{ \ \ \ \ \ }\left[ \hat{p}_{\mu },\hat{p}_{\nu }\right] =0
\end{equation}%
where\textbf{\ I} is the $4\times 4$ identity matrix. Thus the\ $\gamma $'s
satisfy a Clifford algebra and are represented by matrices. Eq.1 is
recognized as the Lorentz invariant Dirac equation with:%
\[
H_{D}=c\mathbf{\alpha .\hat{p}}+\beta m_{0}c^{2},\text{ \ \ }\alpha
^{i}=\gamma ^{0}\gamma ^{i}\text{ \ \ \ }\beta =\gamma ^{0} 
\]

In the same way, canonical quantization\ and factorization of the special
relativity invariant \ $x^{\mu }x_{\mu }=s^{2}:=(\tau _{0}c)^{2}$\ yields a
constraint that is satisfied by the linear equation:%
\begin{equation}
\lbrack \gamma ^{\nu }\hat{x}_{\nu }-\tau _{0}c]\left\vert \Psi
\right\rangle =0
\end{equation}%
provided now that: 
\begin{equation}
\left\{ \gamma ^{\mu },\gamma ^{\nu }\right\} =2\eta ^{\mu \nu }\mathbf{I}%
\text{ \ \ \ \ \ }\left[ \hat{x}^{\mu },\hat{x}^{\nu }\right] =0
\end{equation}%
Eq.3\ is a Lorentz invariant equation satisfied by the self-adjoint time
operator:%
\[
T=\mathbf{\alpha .\hat{r}/}c+\beta \tau _{0}\text{ \ \ \ }\alpha ^{i}=\gamma
^{0}\gamma ^{i}\text{ \ \ \ }\beta =\gamma ^{0} 
\]%
introduced earlier in analogy to the Dirac Hamiltonian\cite{Bauer2}. It
introduces an intrinsic time characteristic $\tau _{0}$\ of the system.

Finally the purely imaginary symmetrized invariant \ $O^{-}:=(\hat{x}^{\mu }%
\hat{p}_{\mu }-\hat{p}_{\mu }\hat{x}^{\mu })$\ \ that satisfies Born's
reciprocity invariance under the transformation \ $\hat{x}^{\nu }\rightarrow 
\hat{p}_{\nu },$ $\hat{p}_{\nu }\rightarrow -\hat{x}^{\nu }$ \cite{Born}
suggests using Planck's constant to accept: 
\begin{equation}
\hat{x}^{0}\hat{p}_{0}-\hat{p}_{0}\hat{x}^{0}=\hat{x}^{i}\hat{p}_{i}-\hat{p}%
_{i}\hat{x}^{i}=i\hbar
\end{equation}%
that insures the satisfaction of the constaint:%
\[
\lbrack \hat{x}^{\mu }\hat{p}_{\mu }-\hat{p}_{\mu }\hat{x}^{\mu }]\left\vert
\Psi \right\rangle =[(\hat{x}^{0}\hat{p}_{0}-\hat{p}_{0}\hat{x}^{0})-(\hat{x}%
^{i}\hat{p}_{i}-\hat{p}_{i}\hat{x}^{i})]\left\vert \Psi \right\rangle =0 
\]%
Eqs.5\ complement the commutation relations \ $\left[ \hat{p}_{\mu },\hat{p}%
_{\nu }\right] =0$ and $\ $\ $\left[ \hat{x}^{\mu },\hat{x}^{\nu }\right] =0$
to yield:\ a) infinite continous range of the four-space and four-momentum
spectra; b) the known configuration\ and momentum representations of the
operators $\hat{x}^{\nu }$\ and\ $\hat{p}_{\nu };$ c) the Fourier transfom
relation between the representations of the system state vector and\ d) the
position-momentum uncertainty relation\cite{Aguillon}.

In the configuration representation, $\Psi (\mathbf{r,}x_{0})=\left\langle
x\mid \Psi \right\rangle $, defining $t:=x_{0}/c$, Eq.1\ reads:%
\begin{equation}
i\hbar \frac{\partial \Psi (\mathbf{r},t)}{\partial t}=\{-i\hbar c\alpha ^{j}%
\frac{\partial }{\partial x^{j}}+\beta m_{0}c^{2}\}\Psi (\mathbf{r},t)
\end{equation}%
recognized as the time dependent Dirac equation for free motion.

In the momentum representation $\Phi (\mathbf{p},p_{0})=\left\langle p\mid
\Psi \right\rangle ,$ defining $e:=cp_{0},$ Eq.3 reads:%
\begin{equation}
i\hbar \frac{\partial \Phi (\mathbf{p},e)}{\partial e}=\{i(\hbar /c)\alpha
_{j}\frac{\partial }{\partial p_{j}}+\beta \tau _{0}\}\Phi (\mathbf{p},e)
\end{equation}%
that clearly relates energy changes to momentum changes.

As a self-adjoint operator, $T$ is the generator of infinitesimal momentum
displacements (Lorentz boosts)\ \ $\delta \mathbf{p}=(\delta e/c)\mathbf{%
\alpha =}(\delta e/c^{2})c\mathbf{\alpha }$\ \ \ and thus indirectly energy
displacements, circumventing Pauli's objection. In the same way, $H_{D}$ is
the generator of infinitesimal space displacements $\delta \mathbf{r}=c%
\mathbf{\alpha (\delta }t\mathbf{)}$\ where $c\mathbf{\alpha =}d\mathbf{r}%
/dt $\ \ is the velocity operator\cite{Thaller}. For a wave packet $%
\left\langle c\mathbf{\alpha }\right\rangle =\mathbf{v}_{gp}$, the group
velocity. Thus $\left\langle \delta \mathbf{p}\right\rangle =(\delta m)%
\mathbf{v}_{gp}=\gamma m_{0}\mathbf{v}_{gp}$ and $\left\langle \delta 
\mathbf{r}\right\rangle =\mathbf{v}_{gp}(\delta t).$

$T$ also generates a phase change $\delta \varphi =\beta (\delta e)\tau
_{0}/\hbar $ while \ $H_{D}$\ \ generates a phase change $\delta \chi =\beta
(\delta t)m_{0}c^{2}/\hbar .$ These are equal provided \ $\delta
e=m_{0}c^{2} $ and \ $\delta t=\tau _{0}$. Furthermore a common finite $2\pi 
$\ phase shift requires $\tau _{0}=h/m_{0}c^{2}$. In conclusion, the
dynamical time operator $T=\mathbf{\alpha .\hat{r}}/c+\beta h/m_{0}c^{2}$\
(where the parameter $\tau _{0}$ is equated to de Broglie period $%
h/m_{0}c^{2}$), generates the Lorentz boost that gives rise to the de
Broglie wave\cite{Baylis}. This also supports the de Broglie period as an
intrinsic property of matter, in agreement with experiment\cite{Lan}.

These Dirac energy and time operators satisfy the commutation relation:%
\begin{equation}
\left[ T,H_{D}\right] =i\{I+2\beta K\}\hbar +2\beta \{\tau
_{0}H_{D}-m_{0}c^{2}T\}
\end{equation}%
where $K=\beta \{2\mathbf{s.l}/\hbar ^{2}+1\}$\cite{Thaller} is a constant
of motion. The related uncertainties are such that $(\Delta T)(\Delta
H_{D})\simeq (\Delta \mathbf{r})(\Delta \mathbf{p})$\cite{Bauer2,Bauer3},
sustaining the interpretation given by Bohr originally: the time uncertainty
in the instant of passage at a certain point is given by the width of the
wave packet which is complementary to the momentum uncertainty and thus to
the energy uncertainty\cite{Bohr}.

In the Heisenberg picture:

\begin{equation}
\frac{dT(t)}{dt}=\frac{1}{i\hslash }[T,H_{D}]=\{\mathbf{I}+2\beta K\}+\frac{2%
}{i\hslash }\beta \{\tau _{0}H_{D}-m_{0}c^{2}T\}
\end{equation}%
that upon integration yields:%
\begin{equation}
T(t)=\{\mathbf{I}+2\beta K+2\beta (1/i\hbar )\tau _{0}H_{D}\}t-(1/i\hbar
)m_{0}c^{2}\beta \int_{0}^{t}dtT
\end{equation}%
as $\hat{H}_{D}$ is constant. The last term introduces an oscillatory
behavior (Zitterbewegung) about a linear time evolution.

The eigenspinors of the self adjoint energy $H_{D}$\ and time $T$\ operators%
\textbf{\ (}Appendix A\textbf{)} provide orthogonal basis \textit{additional}
to the continous vector ones generated by operators $\hat{x}_{\nu }$\ \ and $%
\hat{p}_{\nu }$\footnote{%
This clarifies the confusion addressed by Hilgevoord\cite{Hilgevoord}
between the coordinates of a point in space and the position variables of a
particle ("clearly fostered by the notation $x,y,z$ for both \ concepts");
as well as noting that $t:=p^{0}/c\ $and $H_{D}$ \ are not canonical
conjugate variables.},\ with the following characteristics. The energy
spectrum goes from $-\infty $ \ to $+\infty $, with a $\ 2m_{0}c^{2\text{ \ }%
}$gap at\ the origin. The time spectrum goes from \ $-\infty $ \ to $+\infty 
$, with a $2\tau _{0}$ \ gap at\ the origin. As $\tau _{0}=h/m_{0}c^{2\text{ 
}}$ (the deBroglie or Compton period\cite{Broglie,Baylis}) the gaps are seen
to be complementary. To a small energy gap corresponds a large time gap, and
viceversa.\textit{\ }This complementarity may provide the first indication
that the electron neutrino mass has to differ from zero as this would send
the time gap to infinity. Also in the same way that the energy spectrum
clearly defines non relativistic $(cp\ll m_{0}c^{2})$\ \ and relativistic
energy limits $(cp\gg m_{0}c^{2})$ , the time spectrum recognizes short $%
(r/c\ll \tau _{0};$ $\ r\ll c\tau _{0})$ and long $(r/c\gg \tau _{0};$ $\
r\gg c\tau _{0})$ time or space limits. As in the non relativistic limit
Eq.6 yields the two-component positive energy Schr\"{o}dinger-Pauli equation%
\cite{Thaller}, Eq. 7 results in a two-component positive time short range
approximation.

\section{Second quantization of the energy-momentum and time-space fields}

(\textbf{Note} There are many books on quantum field theory. In this section
the presentation of F. Schawbl, "Advanced Quantum Mechanics", Chapter 13\cite%
{Schwabl} is followed, where the representation of the field is given as a
superposition of free solutions in a finite volume $V$. The passage to
infinite volume is achieved with $\sum_{\mathbf{k}}\left( \frac{m}{Ve_{%
\mathbf{k}}}\right) ^{1/2}\Rightarrow \int \frac{d^{3}k}{(2\pi )^{3}}\frac{%
\sqrt{m}}{k_{0}}$ \ where the factor $\sqrt{m}$ is chosen in order to cancel
the factor $1/\sqrt{m}$ in the spinors, so that the limit $m\rightarrow 0$
exists).

\textbf{a) The time-space representation}

The second quantization of the $\Psi (\mathbf{r},t)$ field considers its
expansion in terms of the spinor eigenvector basis $\{\left\vert
e^{q}\right\rangle \}$ (Appendix A) and transforms the expansion
coefficientes into creation and anhilation (particle and antiparticle)
operators:%
\begin{equation}
\left\langle \mathbf{r}\mid \hat{\Psi}\right\rangle =\hat{\Psi}(\mathbf{r}%
)=\sum_{\mathbf{p},q}\left( \frac{m_{0}c^{2}}{Ve_{p}}\right) ^{1/2}(\hat{b}%
_{q\mathbf{p}}u_{q}^{e_{p}}(p)e^{-i\mathbf{p.r/}\hbar }+\hat{d}_{q\mathbf{p}%
}^{\dagger }w_{q}^{e_{p}}(p)e^{i\mathbf{p.r}/\hbar })
\end{equation}%
where the \textbf{"energy spinors"} are:

\begin{equation}
\left\vert e^{q}\right\rangle =u_{q}^{e_{p}}(p)\left\vert \mathbf{p}%
\right\rangle \text{ \ \ \ }e_{p}>m_{0}c^{2}\text{ \ and\ \ }\left\vert
e^{q}\right\rangle =w_{q}^{e_{p}}(p)\left\vert \mathbf{p}\right\rangle \text{
\ \ \ }e<m_{0}c^{2}-\text{\ }
\end{equation}

\begin{eqnarray}
u_{q}^{e_{p}}(p) &=&\left( \frac{e_{p}+m_{0}c^{2}}{2m_{0}c^{2}}\right)
^{1/2}\left( 
\begin{array}{c}
\chi _{q} \\ 
\frac{c\mathbf{\sigma .p}}{e_{p}+m_{0}c^{2}}\chi _{q}%
\end{array}%
\right) \text{\ } \\
w_{q}^{e_{p}}(p) &=&-\left( \frac{e_{p}+m_{0}c^{2}}{2m_{0}c^{2}}\right)
^{1/2}\left( 
\begin{array}{c}
\frac{c\mathbf{\sigma .p}}{e_{p}+m_{0}c^{2}}\chi _{q} \\ 
\chi _{q}%
\end{array}%
\right)
\end{eqnarray}%
with $\ p=\left\vert \mathbf{p}\right\vert \ ,\ \ e_{p}=+\sqrt{(c\mathbf{p}%
)^{2}+(m_{o}c^{2})^{2}}$ and for $q=1,2\ \ \ \chi _{_{1}}=\left( 
\begin{array}{c}
1 \\ 
0%
\end{array}%
\right) ,$ $\chi _{2}=\left( 
\begin{array}{c}
0 \\ 
1%
\end{array}%
\right) $. Thus $u_{1,2}(p)$ correspond to positive energy and up and down
spin, while $w_{1,2}(p)$\ \ correspond to negative energy and up and down
spin, as can be seen clearly in the rest frame where $\mathbf{p}=0$. The
minus sign in \ $w_{q}^{e_{p}}(p)$\ \ insures that the charge conjugation
operation $C$ transforms the spinors $u_{q}(p)$ into $w_{q}(p)$ and
viceversa.

It then follows:%
\begin{equation}
P^{\mu }=i\int d^{3}x\{\bar{\psi}\gamma _{0}\partial ^{\mu }\psi \}=\sum_{%
\mathbf{p},q}p^{\mu }(\hat{b}_{q\mathbf{p}}^{\dagger }\hat{b}_{q\mathbf{p}}-%
\hat{d}_{q\mathbf{\hat{p}}}\hat{d}_{q\mathbf{p}}^{\dagger })
\end{equation}%
where:%
\begin{equation}
P^{0}=\sum_{\mathbf{p},q}cp_{0}(\hat{b}_{q\mathbf{p}}^{\dagger }\hat{b}_{q%
\mathbf{p}}-\hat{d}_{q\mathbf{\hat{p}}}\hat{d}_{q\mathbf{p}}^{\dagger
})=\sum_{\mathbf{p},q}e_{p}(\hat{b}_{q\mathbf{p}}^{\dagger }\hat{b}_{q%
\mathbf{p}}-\hat{d}_{q\mathbf{\hat{p}}}\hat{d}_{q\mathbf{p}}^{\dagger })
\end{equation}%
Anticommutation of the field operators to satisfy FermiDirac statistics\cite%
{Pauli2} and normal ordering to avoid zero point terms yields:%
\begin{equation}
P^{0}:=H=\sum_{\mathbf{p},q}e_{p}(\hat{b}_{q\mathbf{p}}^{\dagger }\hat{b}_{q%
\mathbf{p}}+\hat{d}_{q\mathbf{p}}^{\dagger }\hat{d}_{q\mathbf{p}})>0
\end{equation}%
i.e., the total energy as the sum of positive \textbf{energy quanta} \ $%
e_{p} $ \ for both states, now interpreted as representing electrons and
positrons respectively. The total momentum is given as:%
\begin{equation}
\mathbf{P}=\sum_{\mathbf{p},q}\mathbf{p}(\hat{b}_{q\mathbf{p}}^{\dagger }%
\hat{b}_{q\mathbf{p}}+\hat{d}_{q\mathbf{p}}^{\dagger }\hat{d}_{q\mathbf{p}})
\end{equation}%
As components of a four vector, $P^{0}$ and $P^{i}$ are part of the
energy-momentum tensor (stress-energy tensor): 
\begin{equation}
T^{\mu \nu }=i\bar{\psi}\gamma ^{\mu }\partial ^{\nu }\psi
\end{equation}%
whose other components yield the total (orbital plus spin) angular momentum
of the system\textbf{.}

The momentum operator $P^{\mu }$\ \ is furthermore shown to be the generator
of space-time displacements, i.e.:%
\begin{equation}
e^{ia_{\mu }P^{\mu }}\Psi (x)e^{-ia_{\mu }P^{\mu }}=\Psi (x+a)
\end{equation}

\textbf{b) The energy-momentum representation}

The above procedure can also be applied to the energy-momentum
representation \ $\Phi (p)=\left\langle p\mid \Psi \right\rangle =\Phi (%
\mathbf{p},p_{0})$ and its adjoint. Expanding the field \ $\Phi (\mathbf{p}%
,E)$\ \ in the spinor $\{\left\vert t_{r}\right\rangle \}$ basis (Appendix
A) yields\footnote{%
Here a finite momentum volume is considered, yielding an unfamliar
discretization of space. However in the infinite volume limit $\ \sum_{%
\mathbf{r}}$ goes to the familiar $\int d\mathbf{r}$.}:

\begin{equation}
\left\langle \mathbf{p}\mid \hat{\Psi}\right\rangle =\hat{\Phi}(\mathbf{p}%
)=\sum_{\mathbf{r},q}\left( \frac{\tau _{0}}{Vt_{r}}\right) ^{1/2}(\hat{a}_{q%
\mathbf{r}}u_{q}^{t_{r}}(r)e^{-i\mathbf{p.r}/\hbar }+\hat{c}_{q\mathbf{r}%
}^{\dagger }w_{q}^{t_{r}}(r)e^{i\mathbf{p.r}/\hbar })
\end{equation}%
with eigenvectors, now \textbf{"time spinors"}:

\begin{eqnarray}
u_{q}^{t_{r}}(r) &=&\left( \frac{t_{r}+\tau _{0}}{2\tau _{0}}\right)
^{1/2}\left( 
\begin{array}{c}
\chi _{q} \\ 
\frac{\mathbf{\sigma .r/}c}{t_{r}+\tau _{0}}\chi _{q}%
\end{array}%
\right) \text{...}t>0 \\
w_{q}^{t_{r}}(r) &=&\left( \frac{t_{r}+\tau _{0}}{2\tau _{0}}\right)
^{1/2}\left( 
\begin{array}{c}
\frac{\mathbf{\sigma .r/}c}{t_{r}+\tau _{0}}\chi _{q} \\ 
\chi _{q}%
\end{array}%
\right) \text{...}t<0
\end{eqnarray}%
with $t_{r}=+\sqrt{(\mathbf{r}/c)^{2}+\tau _{0}^{2}}$ and again $q=1,2$\ ,$\
\chi _{_{1}}=\left( 
\begin{array}{c}
1 \\ 
0%
\end{array}%
\right) $, $\chi _{_{2}}=\left( 
\begin{array}{c}
0 \\ 
1%
\end{array}%
\right) $. Thus $u_{1,2}^{t_{r}}(\mathbf{r})$ correspond to positive time
and up and down spin, while $w_{1,2}^{t_{r}}(\mathbf{r})$\ \ correspond to
negative time and up and down spin.

One now obtains:%
\begin{equation}
T^{\mu }=\sum_{\mathbf{r},q}(x^{\mu }/c)(\hat{a}_{q\mathbf{r}}^{\dagger }%
\hat{a}_{q\mathbf{r}}-\hat{c}_{q\mathbf{r}}\hat{c}_{q\mathbf{r}}^{\dagger })
\end{equation}%
As before, requiring anticommutation and normal ordering yields:%
\begin{equation}
T^{0}=\sum_{\mathbf{r},q}t_{r}(\hat{a}_{q\mathbf{r}}^{\dagger }\hat{a}_{q%
\mathbf{r}}+\hat{c}_{q\mathbf{r}}^{\dagger }\hat{c}_{q\mathbf{r}})>0\text{ \
\ \ }t_{r}=+\sqrt{(\mathbf{r}/c)^{2}+\tau _{0}^{2}}>0
\end{equation}%
i.e., the time operator field contains only positive times. This development
introduces \textbf{time quanta} $t_{r}=+\sqrt{(\mathbf{r}/c)^{2}+\tau
_{0}^{2}}>0$ that are created and destroyed. by the operators $a$ ,$%
a^{\dagger }$ and $c$ ,$c^{\dagger }$.\ In analogy to Eq.17, \ Eq.25\
represents a total intrinsic time associated with the system.

One also obtains from Eq.23\ the vector relation:

\begin{equation}
\mathbf{T}=\sum_{\mathbf{r},q}(\mathbf{r}/c)(\hat{a}_{q\mathbf{r}}^{\dagger }%
\hat{a}_{q\mathbf{r}}+\hat{c}_{q\mathbf{r}}^{\dagger }\hat{c}_{q\mathbf{r}})
\end{equation}%
or$:$%
\begin{equation}
c\mathbf{T}=\sum_{\mathbf{r},q}\mathbf{r}(\hat{a}_{q\mathbf{r}}^{\dagger }%
\hat{a}_{q\mathbf{r}}+\hat{c}_{q\mathbf{r}}^{\dagger }\hat{c}_{q\mathbf{r}})
\end{equation}%
In a similar way $T^{0}$ and $T^{i}$ as components of a four vector are part
of a time-space\ tensor, 
\begin{equation}
\tilde{T}^{\mu \nu }=i\bar{\psi}\gamma ^{\mu }\partial ^{p_{\nu }}\psi
\end{equation}%
\bigskip whose other components yield the space boosts of the system as in
analogy with Eq.20 the time operator $T^{\mu }$\ \ can be shown to be the
generator of energy-momentum displacements , i.e.:%
\begin{equation}
e^{iq_{\mu }T^{\mu }}\Phi (p)e^{-iq_{\mu }T^{\mu }}=\Phi (p+q)
\end{equation}

\section{Interpretation and Conclusion}

The inclusion of the self adjoint time operator and associated
representation restores in QM the equal footing of time and energy accorded
by SR, circumventing Pauli's objection and providing an extended insight on
the problem of time. The time space representation exhibits a total energy
which is the sum of positive energy quanta for both particles and
antiparticles present (Eq.17). Particles with negative energies are
interpreted as antiparticles with positive energies. On the other hand, the
energy-momentum representation exhibits a total intrinsic time as the sum of
positive time quanta for both particles and antiparticles (Eq.25), giving a
formal support to Feynman's extraordinary identification of the negative
energy solutions evolving backward as positrons evolving forward in time.

Second quantization, also referred to as occupation number representation,
is a formulation that allows to describe states with varying numbers of
particles either free or bound by an external or a self consistent
Hartree-Fock potential arising from their interactions. In the each case the
energy quanta are the single particle energies, which comprise the discrete
and continuum energy spectra in the single particle potential $\{e_{i,}e(%
\mathbf{p})\}.$ Then Eq.17 represent the total energy of the independent
particle approximation, where the energy quanta are the single particle
energies. In a description where the ground state is the full Fermi sea,
this energy accounts for the number of paticle-hole (electrons and positrons
or protons and neutrons) states that define an excited state of the system.
In the same way, Eq.25 represents a total intrinsic time summing up the
corresponding individual time quanta expectation values, which are $\pm \tau
_{0\text{ \ }}$for bound single particle states\footnote{%
In the RQM description of atom and nuclear bound states appear as a
consequence of attractive potentials. If these depend solely on position,
e.g., the Coulomb potential in atoms, the shell model self consistent
potential in nuclei, the time operator $T=\mathbf{\alpha .r}/c+\beta \tau _{0%
\text{ }}$satisfies the same commutation relation as with the free particle
Hamiltonian\cite{Bauer2}, The expectation value of $T$ in a bound state $n,.$%
is:%
\[
\left\langle T\right\rangle _{n}=\left\langle n\mid T\mid n\right\rangle
=\left\langle n\mid \mathbf{\alpha .r}/c+\beta \tau _{0\text{ }}\mid
n\right\rangle =\pm \tau _{0\text{ }} 
\]%
as the first term is zero, $\mathbf{\alpha }$\textbf{\ }\ being a non
diagonal matrix.\bigskip}. Thus the intrinsic time increases with the number
of particle-hole pairs involved in an excited state of the system.

An early connection is found in Feshbach's unified theory of nuclear
reactions\cite{Feshbach2,Feshbach} where sharp energy resonances are shown
to be due to compound nucleus states. These are tipically many particle-hole
configurations of mainly single particle bound states in the common
potential. Unbound single particle states connecting to the open reaction
channels will appear with very small amplitudes, leading to long delay times
and long lifetimes\cite{Feshbach,Bauer5,Bauer6}, wheras in the present
formulation they are shown to carry\ large intrinsic times.

To conclude, the purpose of this paper stresses the extension of basic RQM
with the existance a self-ajoint time operator and the additional basis it
provides. Some applications have been allready identified. Observable
effects in experiments that simulate the Dirac equation\cite{Bauer4} and in
tunneling in attosecond optical ionization\cite{Bauer8} have allready been
associated with the existence of the time operator in the first quantization
of SR. The time operator also provides support for the conditional
interpretation of time in QG\cite{Bauer7}. However its full impact in the
extensive development and applications of QFT remains to be explored, both
theoretically and experimentally. To be noted is that Feshbach resonances
are currently subject of extensive reseach in the fields of cold atom systems%
\cite{Chin} (where they provide the essential tool to control the
interaction between atoms), and of Bose-Einstein condensates\cite%
{Calvanese,Chen}. The many facets of the problem of time in QG\cite%
{Anderson,Kuchar} are also to be considered in the present context.

\bigskip

\section{Appendix A \ Time operator eigenvalues and eigenvectors}

A plane wave solution of the form $\Psi (\mathbf{r},x_{0})=e^{-ipx/\hbar
}\psi (\mathbf{r})=e^{-i\{p_{0}x_{0}-\mathbf{p.r}//\hbar \}}\psi (\mathbf{r}%
) $ \ in Eq.8 yields the eigenvalue equation:

\begin{equation}
\{-i\hbar c\alpha ^{i}\frac{\partial }{\partial x^{i}}+\beta
m_{0}c^{2}\}\psi (\mathbf{r})=e\psi (\mathbf{r})  \tag{A.1}
\end{equation}%
with positive and negative eigenvalues:

\begin{equation}
e=\pm \sqrt{(c\mathbf{p})^{2}+(m_{0}c^{2})^{2}}  \tag{A.2}
\end{equation}%
and eigenvectors (\textbf{"energy" spinors}):

\begin{equation}
\left\vert e^{q}\right\rangle =u_{q}(p)\left\vert \mathbf{p}\right\rangle 
\text{ \ \ \ }e>m_{0}c^{2}\text{ \ and\ \ }\left\vert e^{q}\right\rangle
=w_{q}(p)\left\vert \mathbf{p}\right\rangle \text{ \ \ \ }e<-m_{0}c^{2}-%
\text{\ }  \tag{A.3}
\end{equation}%
where%
\begin{eqnarray}
u_{q}(p) &=&\left( \frac{e_{p}+m_{0}c^{2}}{2m_{0}c^{2}}\right) ^{1/2}\left( 
\begin{array}{c}
\chi _{q} \\ 
\frac{c\mathbf{\sigma .p}}{e_{p}+m_{0}c^{2}}\chi _{q}%
\end{array}%
\right) \text{\ }  \nonumber \\
w_{q}(p) &=&-\left( \frac{e_{p}+m_{0}c^{2}}{2m_{0}c^{2}}\right) ^{1/2}\left( 
\begin{array}{c}
\frac{c\mathbf{\sigma .p}}{e_{p}+m_{0}c^{2}}\chi _{q} \\ 
\chi _{q}%
\end{array}%
\right)  \TCItag{A.4}
\end{eqnarray}%
where $\ e_{p}=+\sqrt{(c\mathbf{p})^{2}+(m_{0}c^{2})^{2}}\ \ \ $and$-$for\ $%
q=1,2$ , \ $\chi _{_{1}}=\left( 
\begin{array}{c}
1 \\ 
0%
\end{array}%
\right) $ and $\chi _{_{2}}=\left( 
\begin{array}{c}
0 \\ 
1%
\end{array}%
\right) $. Note that the negative energy spinors have opposite spin
projection to the corresponding positive energy spinors. ("Given this
definition, the charge conjugation operation $C$ transforms the spinors $%
u_{q}(p)$ into $w_{q}(p)$ and viceversa").

\bigskip

For a particle in an attractive $r-$dependent potential with bound states,
these spinors satisfy the equation%
\[
\{-i\hbar c\alpha ^{i}\frac{\partial }{\partial x_{i}}+V(r)+\beta
m_{0}c^{2}\}\psi (\mathbf{r})=e\psi (\mathbf{r}) 
\]%
where $\{e\}=\{E_{i},E\}$ , the $E_{i}^{q}$ being the bound states $\varphi
_{i}^{q}(\mathbf{r})$ eigenvalues and $E^{q}$ the energy continuum of
distorted eigenfunctions.

The expectation value of the time operator in a bound state is then%
\[
\left\langle T\right\rangle _{i}=\int dr\varphi _{i}^{q\ast }(\mathbf{r})\{%
\mathbf{\alpha .r/c}+\beta \tau _{0}\}\varphi _{i}^{q}(\mathbf{r})=\int
dr\varphi _{i}^{q\ast }(\mathbf{r})\beta \tau _{0}\}\varphi _{i}^{q}(\mathbf{%
r})=\pm \tau _{0} 
\]%
as $\mathbf{\alpha .}$is not a diagonal operator in spin space.

In the same way, a plane wave solution of the form $\Phi (\mathbf{p}%
,p_{0})=e^{-ipx/}\phi (\mathbf{p})=e^{-i\{\hat{p}_{0}x_{0}-p.x//\hbar
\}}\phi (\mathbf{p})$ \ in Eq.15\ yields the eigenvalue equation:

\begin{equation}
\{i(\hbar /c)\alpha ^{i}\frac{\partial }{\partial p_{i}}+\beta \tau
_{0}\}\phi (\mathbf{p})=t\phi (\mathbf{p})  \tag{A.5}
\end{equation}%
with positive and negative eigenvalues:

\begin{equation}
t=\pm \sqrt{(\mathbf{r}/c)^{2}+(\tau _{0})^{2}}  \tag{A.6}
\end{equation}%
and eigenvectors (\textbf{"time" spinors}):

\begin{equation}
\left\vert t^{q}\right\rangle =u(r)_{q}\left\vert \mathbf{r}\right\rangle 
\text{ \ \ \ }t>\tau _{0}\text{ \ and\ \ }\left\vert t^{q}\right\rangle
=w(r)\left\vert \mathbf{r}\right\rangle \text{ \ \ \ }t<-\tau _{0}\text{\ } 
\tag{A.7}
\end{equation}%
with

\begin{eqnarray}
u_{q}(r) &=&\left( \frac{t_{r}+\tau _{0}}{2\tau _{0}}\right) ^{1/2}\left( 
\begin{array}{c}
\chi _{q} \\ 
\frac{\mathbf{\sigma .r/}c}{t_{r}+\tau _{0}}\chi _{q}%
\end{array}%
\right)  \nonumber \\
w_{q}(r) &=&-\left( \frac{t_{r}+\tau _{0}}{2\tau _{0}}\right) ^{1/2}\left( 
\begin{array}{c}
\frac{\mathbf{\sigma .r/}c}{t_{r}+\tau _{0}}\chi _{q} \\ 
\chi _{q}%
\end{array}%
\right)  \TCItag{A.8}
\end{eqnarray}%
where \ $t_{r}=+\sqrt{(\mathbf{r}/c)^{2}+(\tau _{0})^{2}}$\ and for $q=1,2$
\ , \ $\chi _{_{1}}=\left( 
\begin{array}{c}
1 \\ 
0%
\end{array}%
\right) $ , \ $\chi _{_{2}}=\left( 
\begin{array}{c}
0 \\ 
1%
\end{array}%
\right) $.

As eigenvectors of self-adjoint energy and time operators, energy and time
spinors constitute complete orthogonal sets $\{\left\vert e^{q}\right\rangle
\},\{\left\vert t^{q}\right\rangle \}$ as: \textbf{(eq. 6.3.15 )}

\begin{eqnarray}
\bar{u}_{q}^{e}(p)u_{s}^{e}(p) &=&\delta _{qs}\text{ \ \ \ }\bar{u}%
_{q}^{e}(p)w_{s}^{e}(p)=0  \nonumber \\
\bar{w}_{q}^{e}(p)\bar{w}e_{s}^{e}(p) &=&-\delta _{qs}\text{ \ \ \ }\bar{w}%
_{q}^{e}(p)u_{s}^{e}(p)=0  \TCItag{A.9}
\end{eqnarray}%
and similar for the time spinors. They both provide representations for a
system state vector.

\end{document}